# Diamond surfaces with lateral gradients for systematic optimization of surface chemistry for relaxometry – A low pressure plasma-based approach

Yuchen Tian[#,1], Ari R. Ortiz Moreno[#,1], Mayeul Chipaux[2,*] Kaiqi Wu[1], Felipe P. Perona Martinez[1], Hoda Shirzad[2], Thamir Hamoh[1], Aldona Mzyk[1], Patrick van Rijn[1], Romana Schirhagl[1,*]

1 Groningen University, University Medical Center Groningen, Antonius Deusinglaan 1, 9713 AW Groningen, Netherlands,

2 Institute of Physics, École Polytechnique Fédérale de Lausanne (EPFL), CH-1015 Lausanne, Switzerland.

*romana.schirhagl@gmail.com

*mayeul.chipaux@epfl.ch

# These two authors contributed equally

**Abstract:** Diamond is increasingly popular because of its unique material properties. Diamond defects called nitrogen vacancy (NV) centers allow measurements with unprecedented sensitivity. However, to achieve ideal sensing performance NV centers need to be within nanometers from the surface and are thus strongly dependent on the local surface chemistry. Several attempts have been made to compare diamond surfaces. However, due to the high price of diamond crystals with shallow NV centers, a limited number of chemical modifications have been studied. Here, we developed a systematic method to investigate a continuity of different local environments with a varying density and nature of surface groups in a single experiment on a single diamond plate. To achieve this goal, we used diamonds with a shallow ensemble of NV centers and introduced a chemical gradient across the surface. More specifically we used air and hydrogen plasma. The gradients were formed by low pressure plasma treatment after masking with a right-angled triangular prism shield. As a result, the surface contained gradually more oxygen/hydrogen towards the open end of the shield. We then performed widefield relaxometry to determine the effect of surface chemistry on the sensing performance. As expected, relaxation times and thus sensing performance indeed varies along the gradient.

**Introduction**

Due to its unique material properties diamond has gained a lot of attention in recent years. Boron doped diamond is widely studied for the fabrication of electronic devices with excellent electrical properties which withstand harsh conditions[1]. Diamond defects are used as stable spin qubits[2,3] but are also used as sensors for magnetic[4,5] or electric fields[6], temperature[7], pressure[8] or the presence of certain chemicals[9,10,11,12]. However, the performance of diamond in these applications is strongly dependent on the surface chemistry[13]. One issue with shallow NV$^-$ centers is the conversion to neutral NV$_0$. This is especially critical for applications where NV center need to be close to the surface. Unfortunately, NV$_0$ does not have the same spin properties as the desired NV$^-$ and thus cannot be used for quantum sensing. Charge transfer to surface traps can be promoted by optical illumination, which leads to conversion of NV$^-$ to less bright NV$^0$ [14,15]. Another important issue is the presence of dangling bonds on the surface which deteriorate the sensing performance of NV centers in their proximity[16,17]. This is especially critical for sensing applications where the diamond defect needs to be within few nanometers from the surface [18]. Differences in the nanoscale environment are also reducing reproducibility of these measurements.

Many different surface treatments have been tested on nanodiamonds, however, nanoscale materials do not necessarily have the same properties as the bulk material and have for instance different reactivity that is often altered by the presence of edges, different phases and irregular shape[19,20].

Several attempts have been made to compare surfaces of bulk diamond. Rosskopf et al. and Ohashi et al. for instance tested fluorine, oxygen, and hydrogen terminated diamond[16,17] but only had one condition (homogeneous distribution of groups) for each termination. Wang et al. investigated electrical properties of boron doped diamond after oxidizing the surface with different methods[21]. Hauf et al. compared the charge states in NV centers after H and O termination[22]. Also, different plasma treatments have been used already to clean diamond, increase NV center creation yield or alter surface or defect properties[23,24,25]. However, single crystal bulk diamond is expensive and thus the conditions that have been investigated so far are very limited (typically not more than 5 conditions are studied per experiment). While all these studies consistently report that surface termination plays an important role, they typically were only able to compare a small number of conditions.

We demonstrate here a more systematic approach which allows to study continuously varying conditions on a single diamond crystal by applying a chemical gradient. While chemical gradients have been produced and applied to other surfaces for other applications[26,27], applying a chemical gradient on a diamond surface has not been done before. To the best of our knowledge this is new for this specific application but also for any other application in diamond. To achieve this goal, we use a method which was developed for the modification of silicone rubber for different applications[28,29,30,31]. We plasma treat the diamond with a shield as shown in Figure 1.

We further investigate the effect of surface chemistry and of roughness on coherence time and thus quantum sensing performance.

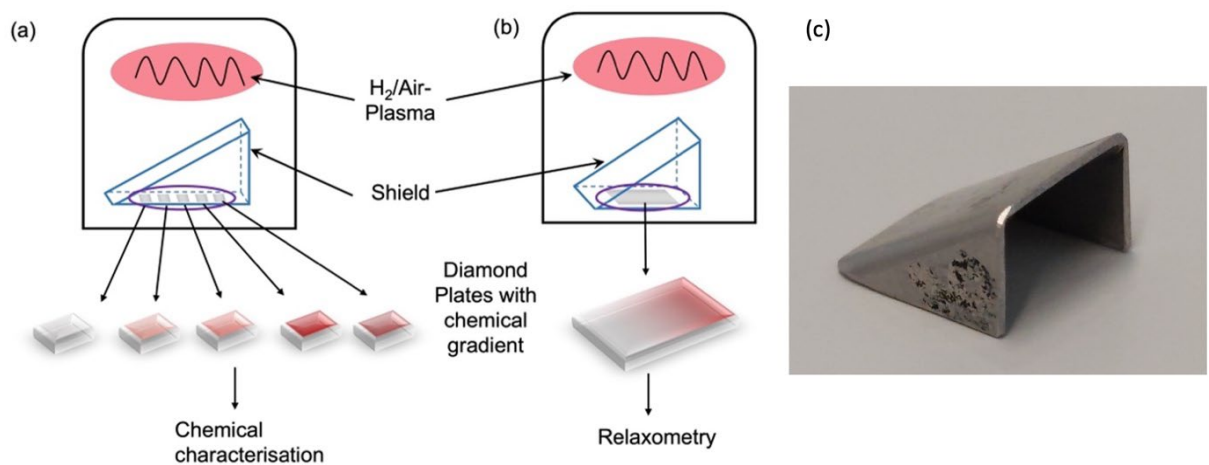

*Figure 1. To generate a chemical gradient across the diamond surface, diamond plates are placed inside a plasma oven. During plasma treatment the surface is partially masked. As a result, the left side of the shield is less exposed than the right.*

*The gradual increase leads to a chemical gradient. This procedure can be done with multiple diamonds (a) or across the surface of a single diamond plate (b). (c) shows a photograph of the triangular prism shield (made of stainless steel) that was used to generate the chemical gradient*

## 2 Materials and Methods

### 2.1. Diamond sample

For surface characterisation we used high pressure-high temperature HPHT diamond surfaces (Element 6). These are cheaper and contain a higher nitrogen content but serve as good samples to estimate surface properties that can be expected in more expensive electronic grade samples with NV centers. The sample treatment in this article is shown in Figure 2.

For relaxometry we used a 2*2*0.5 mm$^3$ electronic grade single crystal diamond plates from Element Six. The diamond was first cleaned using the tri-acid cleaning method as shown before [32,33]. To this end, the diamond is immersed in tri-acid (1:1:1 $HNO_3$:$HClO_4$:$H_2SO_4$) and boiled under reflux in for 1 h (at around 300°C). The sample was cut to expose the [110] face by Almax Easylabs via laser slicing. Then the lasercut facet was polished down to a roughness of Ra<5nm by the manufacturer and this phase was implanted by Cutting Edge Ions. More specifically, $^{15}N^+$. The implantation energy was 5 keV and a dose of 8.27*10$^{12}$ cm$^{-2}$ was used. These conditions create a dense layer of nitrogen and vacancies approximately 8 nm below the diamond surface[34]. Following a previously established protocol the diamond was annealed under vacuum at 800°C for 3 hours. As a result, vacancies combine with nitrogen atoms to form the desired NV$^-$ centers [35].

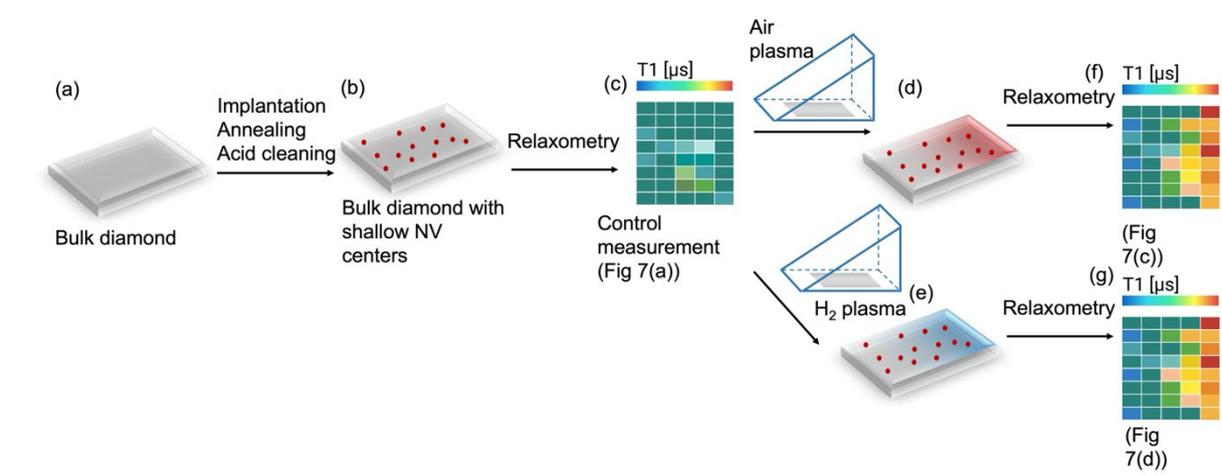

*Figure 2: schematic representation of the experiments in this article (a) We start with a [110] diamond which was implanted, annealed and acid cleaned (b) to obtain shallow NV centers which can be used for quantum sensing in their environment. On these surfaces we performed control measurements (c). These surfaces were then treated with either an air plasma gradient (d) or an H$_2$ plasma gradient (e) These gradients were then further characterised and we collected T$_1$ maps.*

### 2.2. Preparation of Chemical Gradients on Diamond Surfaces.
All the diamond plates were cleaned before every experiment by a well-established acid boiling protocol[33]. Specifically, plates were cleaned in a boiling mixture of sulphuric acid (Sigma-Aldrich, the Netherlands) and nitric acid (Merck, the Netherlands) mixed in the ratio of 1:3 at 140 °C for 4 h. After boiling, the mixture was cooled down to room temperature, then the diamonds were taken out, rinsed multiple times with ultrapure water and subsequently dried with lens cleaning tissue (Thorlabs, Germany). As a result, the surface chemistry before plasma treatment is equal for all samples.

We then performed different kinds of surface treatments using plasma gradients. For the wide field configuration (see § 2.5.1), an air plasma gradient and a hydrogen plasma gradient were applied. The air plasma was generated with 300 W microwave input power (which is the maximum intensity in the chamber) for 30 seconds under stable air pressure of 25 mTorr (Plasma Activate Flecto 10 USB. The hydrogen plasma was created with 800W microwave with 15 mTorr of $H_2$ for 30 s.

In the case of the $\pi$ pulse relaxation times and Hahn Echo measurements performed in a confocal microscope (see 2.5.2), an oxygen plasma treatment (300 W microwaves, 25 mTorr $O_2$, 30 s, Marque) replaced the air plasma treatment. The hydrogen plasma was performed in identical conditions but on a (marque model) instrument.

In every case, as shown in Fig 1, a right-angled prism shield (1 × 1 cm, with an angular aperture of 45° was placed above the diamond surface to induce the desired gradient conditions

## 2.3. Surface characterization

Since chemical surface characterization requires relatively large surfaces this was performed as shown in Fig. 1 (a) on multiple diamond plates that were placed under a shield (or irradiated without a shield as a control).

**2.4.1. Water Contact Angle (WCA) Measurement.** To evaluate the wettability of flat surfaces of diamond plates before and after air plasma treatment, a static WCA measurement was performed using the sessile drop method. To this end we used a homebuilt setup consisting of a camera that uses image analysis to calculate the contact angle. One drop (5 µL) of ultrapure water was placed at the center on each sample. The projected images of the droplets, after having been settled on the substrates with no noticeable change in their shapes, were analyzed for determining contact angles.

For hydrogen plasma treated samples we performed measurements of dynamic contact angles due to the smaller differences in wettability and because these samples were prepared in a different laboratory where this equipment is available. For these surfaces contact angle measurements were performed using Teflon coated glass capillaries for defined deposition and redispersion of small amounts of water (8 µL). We recorded advancing angles (an average of the tracked angle during expansion of the droplet) and the receding angle (an average of the tracked angle during shrinking). Both were measured during a moving three-phase line/wetting front. We repeated these measurements four times on the same spot after drying with nitrogen. The shown data points are the mean and deviation of these four measurements.

The "Post-treatment" conditions refer to samples left in ambient air for one day after the hydrogen gradient was applied.

**2.4.2 X-ray Photoelectron Spectroscopy (XPS).** The air plasma treated diamond plates were characterized using an XPS (S-Probe, Surface Science Instruments, Mountain View, CA, United States) equipped with an aluminum anode. Samples were placed in the prevacuum chamber of the XPS and then subjected to a vacuum of 10−9 Pa. X-rays (10 kV, 22 mA) at a spot size of 250 × 1000 µm were produced using an aluminum anode. Scans of the overall spectrum in the binding energy range of 0–1100 eV were made at low resolution (pass energy 150 eV). Since hydrogen plasma treated samples were fabricated in a different location they were analysed freshly with a different instrument.

**2.4.3. Scanning Electron Microscopy (SEM).**

In preparation for SEM the diamond samples were sputter coated with a 30 nm gold layer using a Balzers SCD050 sputter coater. SEM images were collected using a Philips XL30S SEM FEG instrument. We used secondary electron detection at a working distance of 9.9 mm and a beam energy of 5 keV.

**2.4.4. Atomic Force Microscopy (AFM).** AFM images were obtained using a commercial atomic force microscope (Nanoscope V Dimension 3100 microscope, Bruker, United States) operated in contact mode in air, using a NP-D tip from Bruker. The roughness (Ra) of plates before and after plasma treatment was analyzed based on AFM images using the NanoScope Analysis software.

**2.4. Stability of the gradient**

After plasma treatment, the plates were stored under different conditions to prevent them from reacting in the air and to determine the cause of instability of the surface chemistry. The approach was based on Zhou et al.'s work[36]. There were three diamond plates which were treated by air plasma. Then the plates were stored under air, nitrogen, or water, separately for 7 days. The contact angles of diamond plates were recorded each day.

**2.5 Relaxation and coherence time measurement**

**2.5.1 All optical relaxation time sequences (wide field configuration)**

The aim of relaxometry measurements was to determine the suitability of a certain surface chemistry along the chemical gradient for relaxometry measurements. Relaxometry measurements were performed on a single diamond plate as shown in Fig. 1 (b)
The diamond sample was analyzed with a home built wide-field relaxometer (a widefield microscope with epifluorescence excitation and specific pulsing capabilities) described in a previous work[37]. In this instrument, light excitation is directed onto the bulk diamond from the top and the NV photoluminescence is collected at the bottom[37]. The excitation source is a laser diode module (Coherent Dilas 520 nm), modified to contain only two of the 3 internal laser diodes of the standard model to comply with the driver electrical characteristic. It has been coupled with a laser diode driver from Analog Modules, Model 762. The pulsing of the laser diode is thus performed using the direct modulation of the diode current. An Aardvark I2C/SPI Host Adapter is used to link the computer with the Analog Modules driver. A camera, Andor Zyla, is used to acquire an image of the surface. Collimation of the laser diode is obtained using a fibre collimation package (Thorlabs F230SMA). Further correction is achieved using two lenses. The collimated beam is focused onto the diamond surface using a 15-mm focal lens. The photo-luminescence (PL) of the NV-centres is collected using a 50X microscope objective, a 175 mm focal field lens and a 600 nm fluorescence long-pass filter in front of the camera.
The wide-field microscope was controlled using a custom Labview program, providing in each relaxometry experiment a set of images encoded with their corresponding dark time.

The $T_1$ relaxation measurements provide information about the surrounding of the NV centers and is suitable for detecting magnetic noise[38]. In a typical $T_1$ measurement, the NV centers are excited and thus pumped into the ms=0 state of the ground state. After varying dark times, we probe whether the NV centers are still in this state or have returned to the equilibrium between ms=0 and ms=+-1. This can be concluded from the fluorescence brightness. This process is faster in presence of spin noise. Since spin noise on the surface is deteriorating the sensing performance, a larger $T_1$ can be used as a quality criterion for the local surface chemistry of the diamond. However, in our instrument the dark counts are not exactly zero. To compensate for this, we performed two sets of measurements, one at high power and the other one at low power (as reference). Then we computed the difference between the two. From this procedure, we obtained the $T_1$ relaxation curve per pixel.

After obtaining the photoluminescence vs dark time curve (per pixel), a single exponential fit was performed per pixel and across the whole image.

### 2.5.2 Microwave relaxation time Hahn Echo measurements (confocal configuration)

Microwave $T_1$ and $T_2$ coherence time measurements were performed on another homemade setup, in a confocal microscopy configuration. Laser pulses (515 nm, 2 mW, 100 µs) were focused on a long working distance microscope objective (Olympus X50, $NA = 0.5$). The setup and its interface with the participative Python library QUDI [39], is fully described in our previous publication [40]. A magnetic field bias of a few militesla was applied to distinguish the four NV center groups (according to each crystallographic orientation). One group (at around 2.84 GHz) was selected. At first, Rabi cycles were performed (Fig. S5) to determine the optimal $\pi$ and $\pi/2$ pulse durations. As in our previous work [41] microwave relaxometry sequences, in which an additional microwave $\pi$ pulse is added or omitted right before the next laser pulse, were performed to exclude the spin insensitive processes. 50 laser pulses were logarithmically spanned from 1 µs to 10 ms. Finally, Hahn Echo sequences measuring the coherence time ($T_2$) where performed. The resulting $T_1$ and $T_2$ times are shown in Figure 9. The raw relaxometry and Hanh Echo curves are shown in Figure S5.

## 3. Results and Discussions

### 3.1. Gradient characterization

Generally, a plasma can influence a diamond surface in two ways: etching and chemical modification[42,43,44,45]. The conditions for plasma treatment (Plasma Activate Flecto 10 USB, maximum intensity, 30s) were chosen to avoid etching. The assumed mechanism is that the surface modification effect of a low pressure (25 mTorr and for 30 seconds) plasma treatment is due to the impact of (1) activated particle species like ions, radicals, etc. and (2) high energetic vacuum ultraviolet photons. While (1) is gradually reduced in a more and more narrow gap, (2) is fully blocked by this kind of shield. More aggressive settings could also be used to remove material and bring NV centers closer to the surface or to increase roughness deliberately. Here we will first discuss the impact of plasma treatment on the chemical characteristics of the surface and then morphology and roughness.

The first step we took here was to confirm that we indeed created a chemical gradient on a diamond surface. While this has been observed on several other materials, this has not been shown for diamond yet[46,47,48]. A simple and straight forward way to do this is to use the water contact angle, which indicates changes in hydrophilicity. Figure 3 shows the water contact angle depending on the position under the shield.

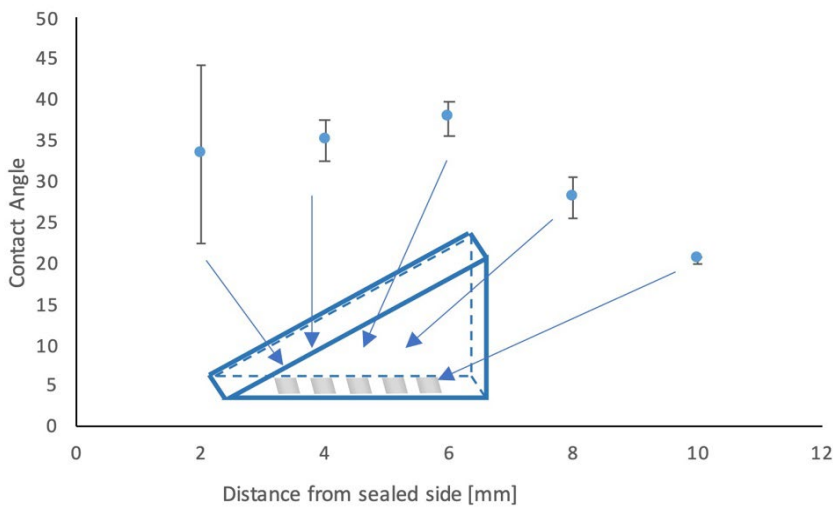

*Figure 3. Formation of a chemical gradient during plasma treatment. Right after treatment with air plasma, the contact angle decreases depending on the position under the shield.*

As expected, we observed the greatest change in contact angle close to the shield opening. Towards the closed end, the surfaces were less affected by the plasma.

To obtain more details on the changes in surface chemistry we performed XPS measurements. This method, which has already been used successfully to assess diamond surface chemistry by us and others, was chosen for its relative simplicity and specificity for the surface[33,49,50]. Figure S2 – S4 show the results from these measurements. We compared three different spectra. The first spectrum is from the untreated diamond surface (shown in Figure S2). The second and third are collected after applying the plasma treatment and forming a gradient. The second (shown in Figure S3) is from the side that was shielded more and thus received less plasma. The third (shown in Figure S4) is from the exposed side of the sample. As expected from an air plasma, the plasma treatment increases the oxygen content and the higher oxidation levels increase in abundance after plasma treatment. The fact that this is more pronounced for the more exposed side confirms that we indeed created the desired chemical gradient across the surface. Even though the starting material is already an oxygen terminated diamond, the plasma indeed further increases the oxygen content on the surface. It is also worth mentioning that we see a contamination peak of Si. This might be due to some surface contamination or intrinsic Si. XPS data for the hydrogen plasma treated surfaces where we confirmed that the plasma led to a change in surface chemistry are shown in Supplementary Figure S1. Further in Table S1 we analyze the C1s peak in more detail. For all the samples the largest contribution to the peak is the carbon which is part of the diamond lattice. For the oxygen plasma, we also here see an increase in peaks that indicate carbon that is bound to oxygen. In addition, there is a peak that could be either SP2 carbon or CH related. For the oxygen plasma we observe a slight decrease in this peak, which is expected. Under hydrogen plasma we see an increase in this peak which might be due to graphitization or due to an increase in CH bonds or a combination of both.

To exclude that the plasma changes the surface roughness or morphology significantly here we performed SEM and AFM imaging. The results are shown in Figure 4.

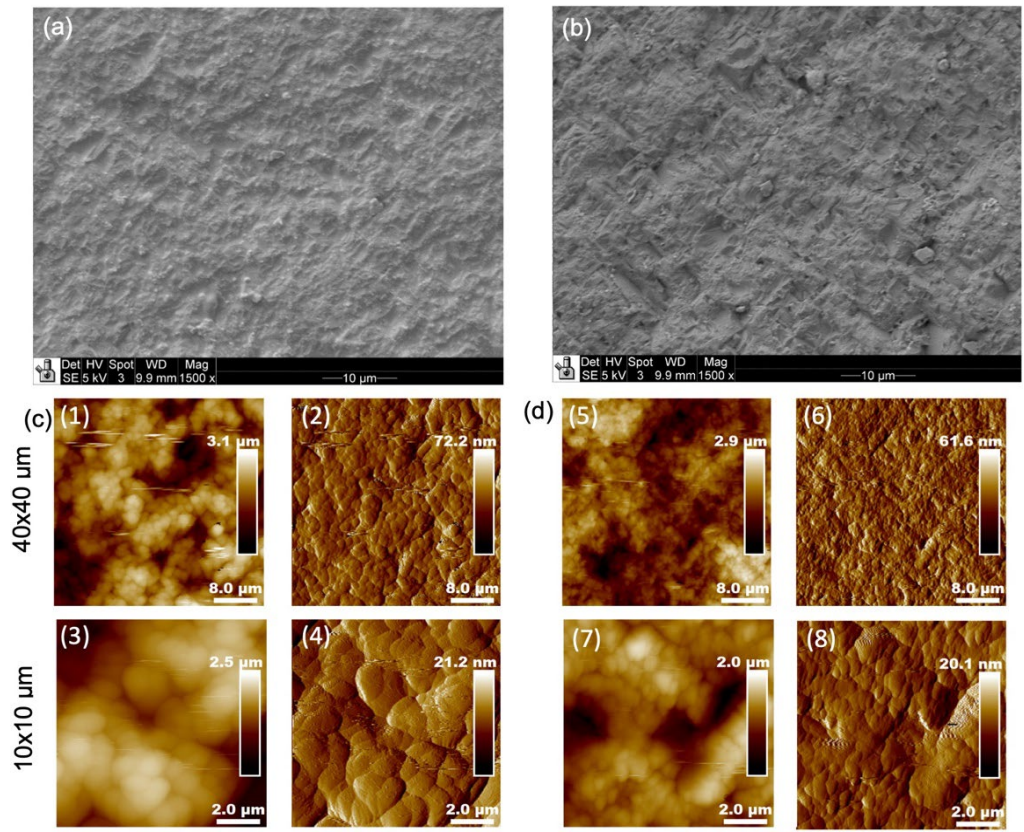

*Figure 4. Analysis of morphology and roughness before and after air plasma treatment (a) shows and SEM image of the diamond surface before and (b) after treatment. (c) and (d) are representative images of the surfaces before and after treatment, which were used to assess the surface roughness. ((1) and (2) as well as (3) and (4) show the same area in height and deflection mode of the sample before treatment and (5) and (6) and (7) and (8) are the same areas in height and deflection mode but the images were taken after exposure to air plasma)*

As shown in Figure 4 (a) and (b) there are no obvious differences in morphology of the diamond surfaces before and after plasma treatment. In addition to these measurements we performed AFM measurements due to the greater sensitivity in z. A more quantitative picture of the roughness can also be obtained by AFM shown in Figure 4 (c) and (d). When analysing the roughness, we revealed a slight increase in roughness caused by the treatment after air plasma treatment (342±13 nm for the pristine and 390±1.4 nm for the plasma treated samples).

### 3.2. Stability of the gradient

After characterizing the gradient, we investigated the stability of the gradient. To this end we monitored the water contact angle over the course of a week. While we observed a drop in the contact angle after treatment, the contact angle slowly returns to its original value in about 3 days. The results of these measurements are shown in Figure 5 (a-c). A similar behavior has been observed by Zhou et al. in silicone gradients where storage in different environment would inhibit (in water and PBS) or promote (in air) hydrophobic recovery of wettability gradients [51]. There are multiple reasons which might explain this behavior including reactions of surface groups in air, adsorption of organic material from the atmosphere or rearrangements of groups on the surface. To determine the cause of instability, we also conducted experiments where the plates were stored differently. That the process also occurs in water or under nitrogen atmosphere indicates that it is independent of air oxygen as well as dust or molecules from the air settling on the plates.

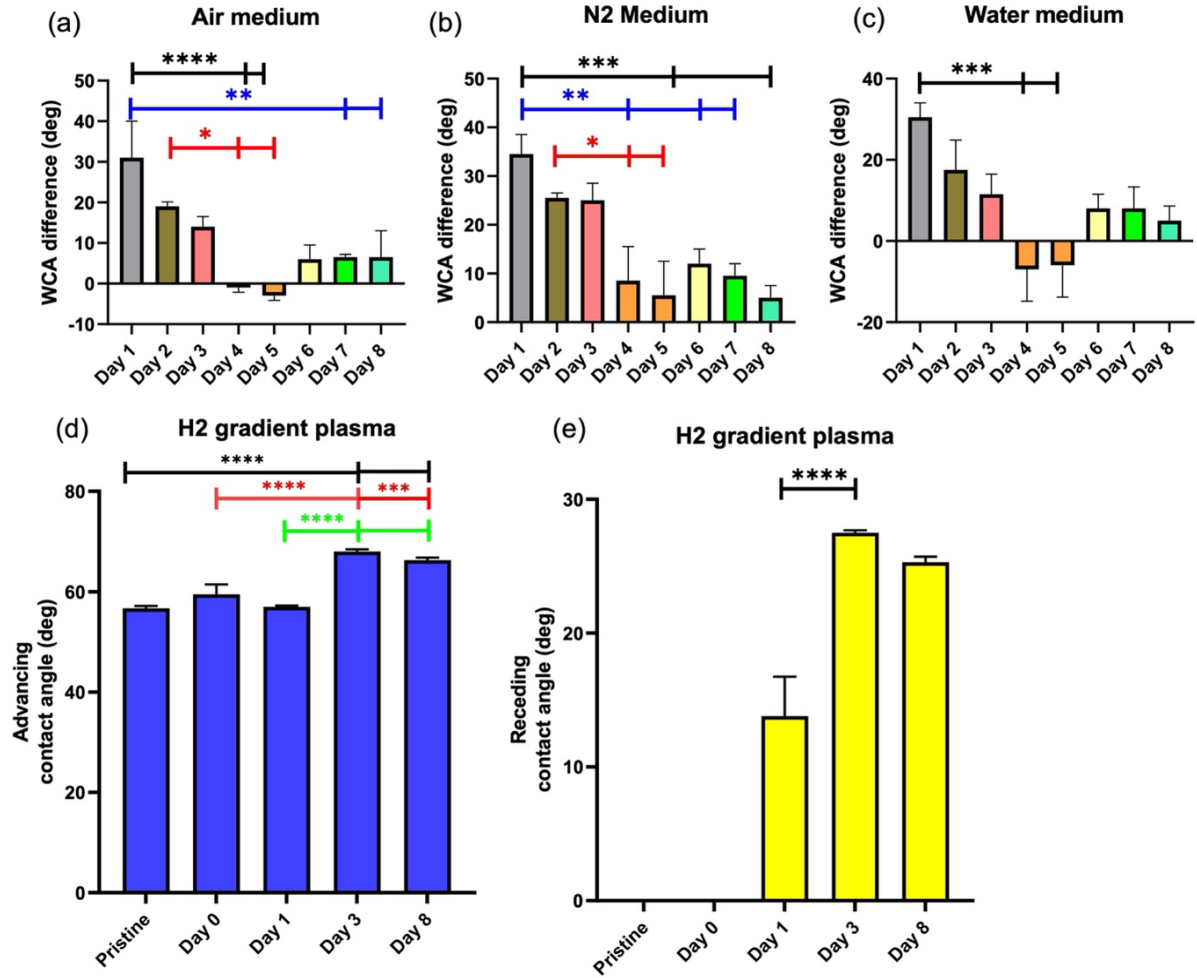

*Figure 5. Stability of the surface modification under different conditions. The data in (a-c) shows water contact angles of air plasma treated surfaces over the course of 8 days in (a) air, (b) $N_2$ atmosphere and (c) water. (d) and (e) show the advancing and receding contact angles. Each experiment was performed 6 times. Statistical significance was determined using an ANOVA, * p < 0.05, ** p < 0.01, *** p < 0.001, **** p < 0.0001*

Also, for the hydrogen plasma treatment we determined the stability of the surface termination by contact angle measurements (Figure 5 (d) and (e)). The dynamic measurement of advancing and receding contact angle gives more information than the static one. The mean of advancing and receding angle can be considered as the mean hydrophilicity. The difference between both reveals topographical and chemical homogeneity [52]. Directly after the hydrogen plasma treatment the wetting of the surface was not much different compared with the pristine state. One day later the advancing angle remained the same, while the receding was not zero anymore. After three days the surface became clearly more hydrophobic, but this normalized again over two weeks, showing a trend back to conditions of the pristine surface. We interpret this dynamic change as a chemical reorganization, first, from a hydrophilic surface to a chemically more homogeneous and hydrophobic condition.

### 3.3. Relaxation Time Measurement

Finally, we assessed the influence of the surface modification on the sensing capabilities of the NV centers below the surface. There are multiple ways to do diamond magnetometry measurements, but they have in common that they are limited the relaxation time ($T_1$) or coherence time ($T_2$) [13,53,54,55]. These times indicate how long an NV$^-$ center can remain in a prepared state before returning to equilibrium or how long the quantum coherence can be maintained (in the Bloch sphere equatorial plane). Thus, they determine the quality of the NV⁻ centers and their sensing performance. First, we characterize the $T_1$ (see Fig. 6 to 8) particularly relevant for sensing of free radicals [56]. This sensing scheme does not require microwave excitation and is thus attractive for biomedical applications. These measurements were performed in a widefield-configuration to assess the entire surfaces at once (see 2.5.1). With an additional microwave field applied, we also characterize the microwave $T_1$, excluding spin insensitive processes, and Hahn Echo $T_2$ coherence times (see Fig. 9).

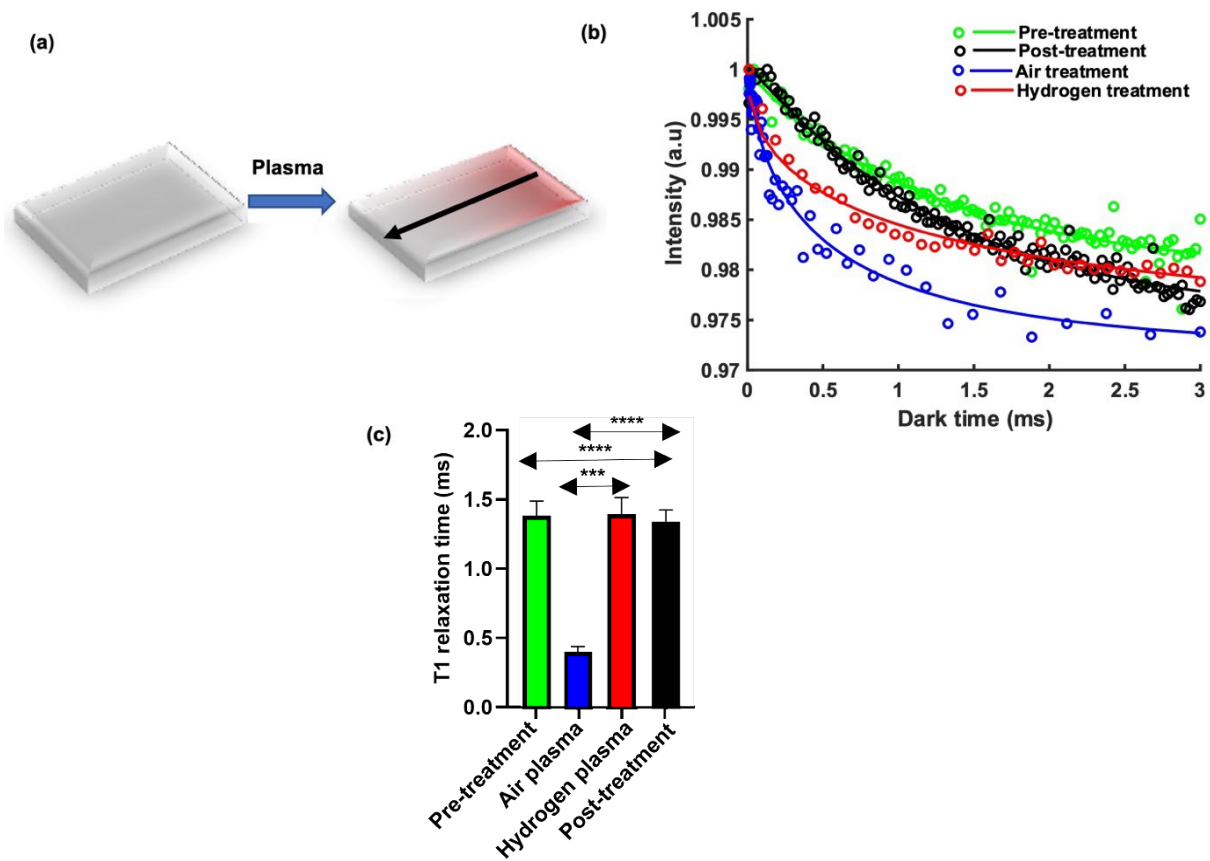

*Figure 6. Relaxometry of the diamond surfaces. (a) Schematic of the combined effect of the shield and the plasma here we investigate the average surface while we investigate the gradient in Figure 7. (b) $T_1$ curves averaged from the entire diamond surface for each condition c) The mean $T_1$ values that were extracted from $T_1$ curves from the entire surface. The pre-treatment group refers to a condition where the surface has been acid cleaned while for post treatment the sample has been left in ambient air for one day after treatment with hydrogen plasma. Each experiment was performed 5 times. \*\*\* p < 0.001, \*\*\*\* p < 0.0001*

3.3.1. All optical, widefield, measurements

While we calculate $T_1$ in nanodiamonds with a double exponential fit[57],[58] we here used a single exponential model. While the single exponential poorly fitted the data for nanodiamonds, with bulk diamonds we did not have that problem. NV centers in bulk diamonds are more uniform and thus can be described sufficiently by just one exponential. There are several possible explanations for having less uniform NV centers in nanodiamonds: 1) With implantation the depth where NV centers occur is more uniform. 2) there is only 1 surface that is close enough to have an influence on the coherence time/relaxation time in bulk and it is the same plane for all the NV centers (while in nanodiamonds there might be some NVs that are close to a 111 surface while others are close to 110 or 100). 3) In bulk, edges are negligibly far away while in nanodiamonds, edges probably play a large role. 4) the nanodiamonds we used are HPHT in origin and thus have more impurities.

From Figure 6 we can observe several trends. First, air plasma treatment reduces $T_1$ significantly. This can be explained by dangling bonds which are introduced during plasma treatment. While they also report a shortening in coherence times after oxidative etching this is different from the effect that Wang et al. have observed[59]. In their article they used hours of air oxidation. Under such conditions, the diamond is etched substantially and the changes in coherence times can be attributed to the closer proximity to the surface. Also, Kim et al. used plasma treatment and its effect on coherence times but in combination with heat treatment[23]. This is also different from the improved coherence times after plasma treatment that have been observed by Osterkamp et al. who used $SF_6$ plasma which leads to a surface which is chemically very different[60].

In contrast, hydrogen plasma treatment changes the average $T_1$ of the surface only slightly (insignificantly compared to the acid regenerated surface and significantly compared to oxygen plasma treated surfaces). This is a somewhat surprising finding since there are some cases where hydrogen termination has been shown to be worse than oxygen termination[61]. In our case however, an air gradient (instead of an oxygen gradient) is applied which might result in a different surface chemistry (see also 3.3.2).

Another possible explanation might be that there are different levels of graphitisation. However, from the samples after air plasma, we see a decrease in the peak which might indicate graphitisation in XPS. For the hydrogen plasma we see an increase in the respective C1s peak in the XPS data (Table S1). However, this peak likely also contains a contribution from CH and thus doesn't give a direct measure for the graphitisation.

A further concern is if the measurements in the different surfaces were performed on the same NV centers. It has been reported before that oxygen termination stabilizes NV centers that are closer (which regardless of the surface chemistry have a lower $T_1$ [62]) to the surface while hydrogen termination destabilizes NV centers that are closer to the surface[63]. We expect this effect to be reduced by the fact that we implant nitrogen in a specific depth. In the light of this point we have added measurements of the photoluminescence to the manuscript (see Figure S6). If more NV centers close to the surface would be stabilised, we would expect a higher count rate. However, we do not see an increase in counts (which would indicate that more NV centers contribute). The only change we have observed is relatively low counts at the opening where the highest dose of the plasma was received. This effect can probably be attributed to slight etching of the surface.

That we did not observe a drastic change in overall $T_1$ after hydrogen plasma treatment could mean that the surface does not change at all or that there are areas where the $T_1$ improves and areas where it decreases. To differentiate the two, we further investigated $T_1$ maps.

The $T_1$ maps of the different surfaces are shown in Figure 7. Here we can clearly see that the surface before treatment (Figure 7 (a,c)) is not very uniform in terms of $T_1$ values over the surface. A potential reason might be dangling bonds or adhering molecules on the surface which are affected by the surface treatment. When we applied an air plasma gradient (Figure 7 (b,e)), the relaxation times decrease gradually from the least to the most exposed area of the sample. It is also clear from these data that the entire surface has lower $T_1$ values than before treatment. For hydrogen plasma treatment (Figure

7 (d) and Figure 8 (d)) we also observe a gradient. However, in this case we did not observe the drastic shortening of $T_1$ over the entire surface. Rather we see that part of the sample has increased $T_1$ while the other parts of the sample have lower $T_1$ (potentially from etching). What is most interesting about this approach is that we were able to investigate different plasma conditions which change the surface chemistry continuously in a single experiment with only one diamond. Here it also has to be noted that we are currently limited by the surface area that we can observe due to the small size of the diamond as well as the field of view of the setup. While our sample had a densely packed ensemble, we believe that similar trends would also apply to other diamond substrates. If a sample with single NV centers or higher quality of diamond, we would expect a similar trend but at overall higher $T_1$ values.

We further observed the $T_1$ for surfaces that were treated with hydrogen plasma again after waiting for one more day. Consistent with the results for the contact angle measurements and with the results for the air plasma, we also here observed that the changes in $T_1$ are not permanent but the original $T_1$ is restored.

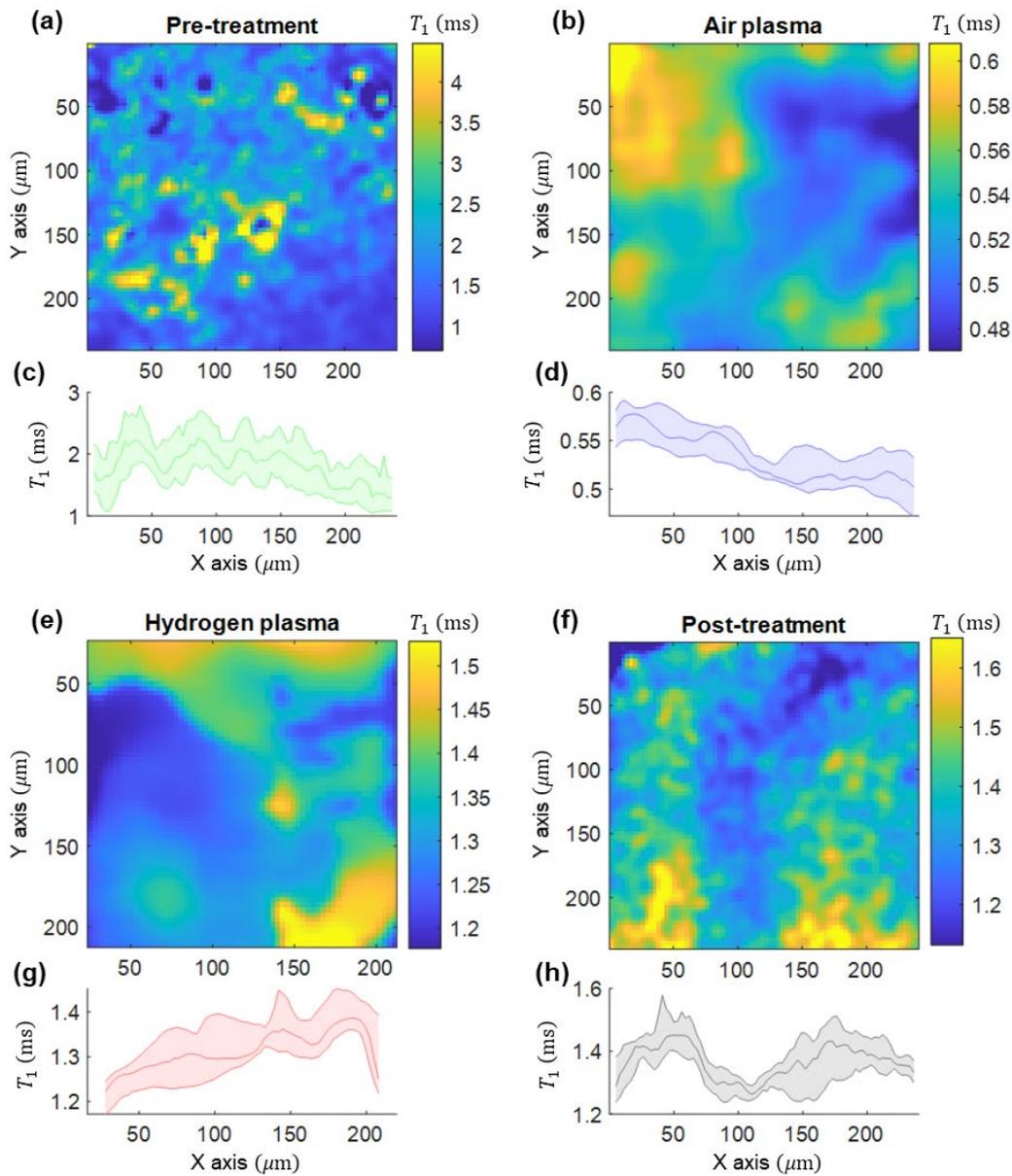

*Figure 7. Relaxometry maps (a, b, e, f) of the diamond surface for different conditions. The gradients were applied along the x-axis. The average (c, d, g, h) over each column (y axis) with the median and the interquartile range are shown to better visualize the gradient. (a,c) Pre-treatment (before any plasma is applied) (b,d) After an air gradient plasma treatment (e,g) After a hydrogen gradient plasma treatment. (f,h) Post-treatment, after recovery (one day after hydrogen plasma treatment).*

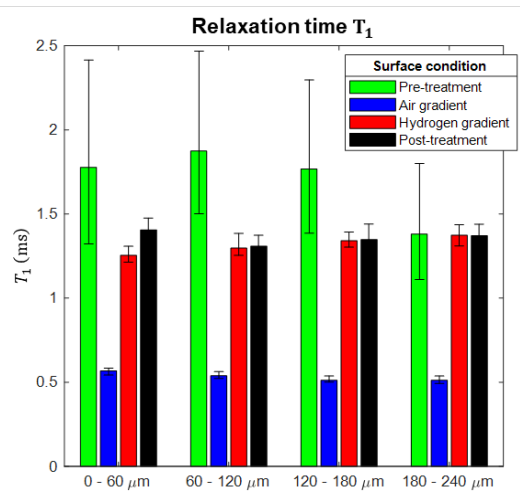

*Figure 8. Comparison of $T_1$ resulting from the different surface treatments. Each $T_1$ image was divided in four areas. The bars represent the median of each subdivision. The error bars correspond to the interquartile range.*

3.3.2. All optical, widefield, measurements

At first, the oxygen plasma significantly increased the relaxation $T_1$, up to about 5 ms, which appeared to be uniform over the surface. Notably, as opposed to the effect of the other treatment, this effect is long lasting. It also appeared uniform. As opposed to the air plasma, the oxygen plasma therefore behaved as expected from literature where oxygen treatment has been shown to reduce dangling bounds[23], saturating already after a short/low energy exposure. The hydrogen plasma then decreased the relaxation time down to the values observed in wide field. As shown in Figure 9, that effect is however clearly influenced by the gradient, in the opposed direction from the wide field condition which were preceded by a slightly decreased $T_1$ (instead of increased) induced by the air plasma.

Oppositely, we did not find any significant change in the $T_2$ times (of a few microseconds comparable with values in the literature[64]) measured through Hahn echo sequences (see Fig S5). $T_2$ is indeed expected to be more influenced by the internal nitrogen concentration than by the surface[65].

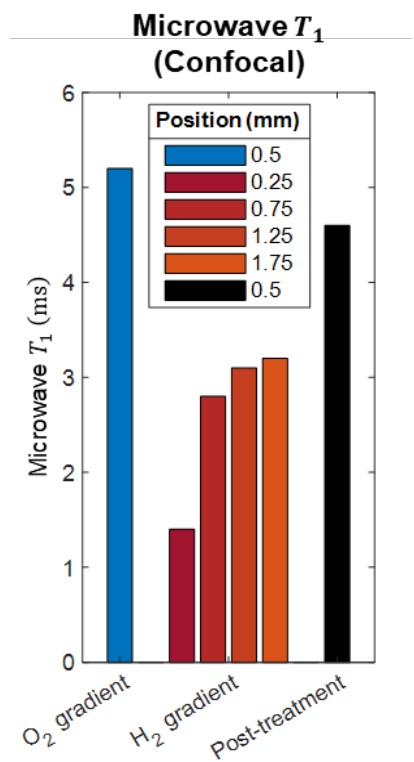

*Figure 9. Microwave $T_1$ after oxygen plasma treatment, hydrogen plasma treatment on four positions along the axis x, and the Post-treatment condition.*

**Conclusions**

Due to the high price of diamond plates, it is usually not feasible to investigate more than a handful of surface conditions. We have demonstrated for the first time here that it is possible to form a chemical gradient on a diamond surface. Such a gradient approach here, which allows to investigate different conditions varying continuously on a single plate. We also showed that the $T_1$ relaxation time changes based on the position along the gradient. While we did not observed improvements in $T_1$, we are confident that this high throughput screening approach is a powerful tool to optimize diamond surfaces. A side-finding of this work is on the stability of the modification. While changes in surface chemistry are often assumed to be permanent, it is important to keep in mind that they might not be infinitely stable. As we observed here, both hydrogen plasma treatment as well as air plasma treatment was only stable over a few days.

**Acknowledgements**

We would like to thank the Dutch Research Council (NWO) for financial support via a VIDI grant (016.Vidi.189.002) and the Swiss National Science Foundation (SNSF) via an Ambizione grant (No. 185824).

---

1 Marselli, B., Garcia-Gomez, J., Michaud, P.A., Rodrigo, M.A. and Comninellis, C., 2003. Electrogeneration of hydroxyl radicals on boron-doped diamond electrodes. Journal of the Electrochemical Society, 150(3), p.D79.
2 Levine, E.V., Turner, M.J., Kehayias, P., Hart, C.A., Langellier, N., Trubko, R., Glenn, D.R., Fu, R.R. and Walsworth, R.L., 2019. Principles and techniques of the quantum diamond microscope. Nanophotonics, 8(11), pp.1945-1973.

# Supplementary Material

Diamond surfaces with lateral gradients for systematic optimization of surface chemistry for relaxometry – A low pressure plasma-based approach

Yuchen Tian[#,1], Ari R. Ortiz Moreno[#,1], Mayeul Chipaux[2,*] Kaiqi Wu[1], Felipe P. Perona Martinez[1], Hoda Shirzad[2], Thamir Hamoh[1], Aldona Mzyk[1], Patrick van Rijn[1], Romana Schirhagl[1,*]

1 Groningen University, University Medical Center Groningen, Antonius Deusinglaan 1, 9713 AW Groningen, Netherlands,

2 Institute of Physics, École Polytechnique Fédérale de Lausanne (EPFL), CH-1015 Lausanne, Switzerland.

*romana.schirhagl@gmail.com

*mayeul.chipaux@epfl.ch

Diamond surfaces with lateral gradients for systematic optimization of nanoscale MRI – A low pressure plasma-based approach

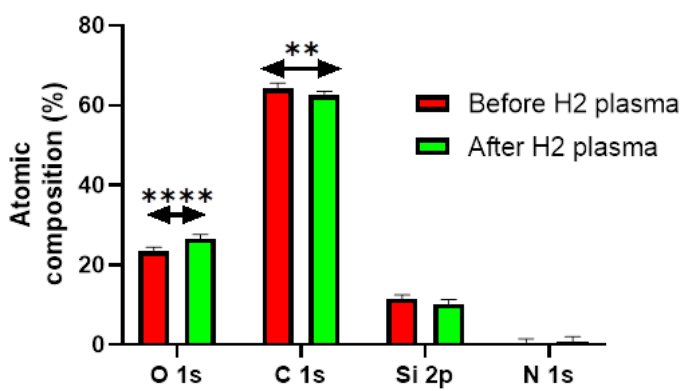

Figure S1: Changes in the atomic composition of the diamond surfaces due to hydrogen plasma treatment. Statistical significance was determined using an ANOVA *** $p < 0.001$, **** $p < 0.0001$. The slight increase in oxygen concentration might occur after the treatment itself when the sample is exposed to ambient air.

Table S1: Analysis of the C1s peak in XPS. Peak assignments are done based on [1,2]

| Diamond before plasma | Position | Area | % | Assignment |
|---|---|---|---|---|
| | 284,8 | 22957 | 13,5 | SP2/CH |
| | 286,53 | 145101 | 85,32 | SP3 |
| | 288,38 | 1437 | 0,84 | CO |
| | 289,74 | 572 | 0,34 | CO |

| Air plasma (less exposed) | Position | Area | % | Assignment |
|---|---|---|---|---|
| | 284,8 | 12576 | 10 | SP2/CH |
| | 286,41 | 107015 | 85,08 | SP3 |
| | 288,16 | 5078 | 4,04 | CO |
| | 289,95 | 1108 | 0,88 | CO |

| Air plasma (more exposed) | Position | Area | % | Assignment |
|---|---|---|---|---|
| | 284,8 | 19674 | 11,93 | SP2/CH |
| | 286,45 | 137074 | 83,11 | SP3 |
| | 288,12 | 8175 | 4,96 | CO |

| H2 plasma | Position | Area | % | Assignment |
|---|---|---|---|---|
| | 284,9707 | 8465,39 | 24,03 | SP2/CH |
| | 285,9386 | 24784,04 | 70,34 | SP3 |
| | 287,1296 | 1984,59 | 5,633 | CO |

XPS data from air plasma treated samples

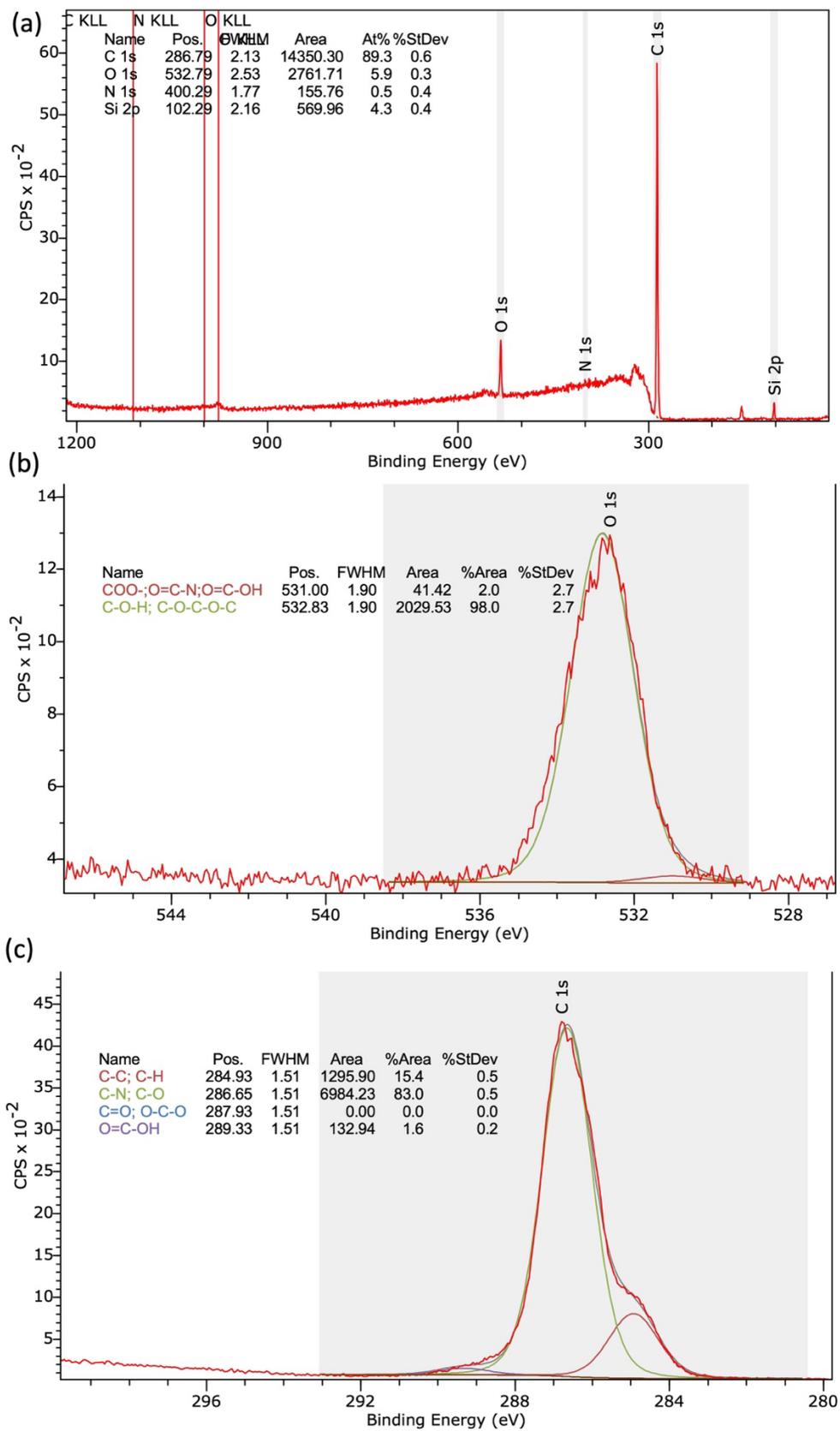

Figure S2. Characterisation of the diamond surface before plasma treatment (a) shows the entire XPS spectrum (b) an analysis of the oxygen group peak. (c) is the area of he C1s peak

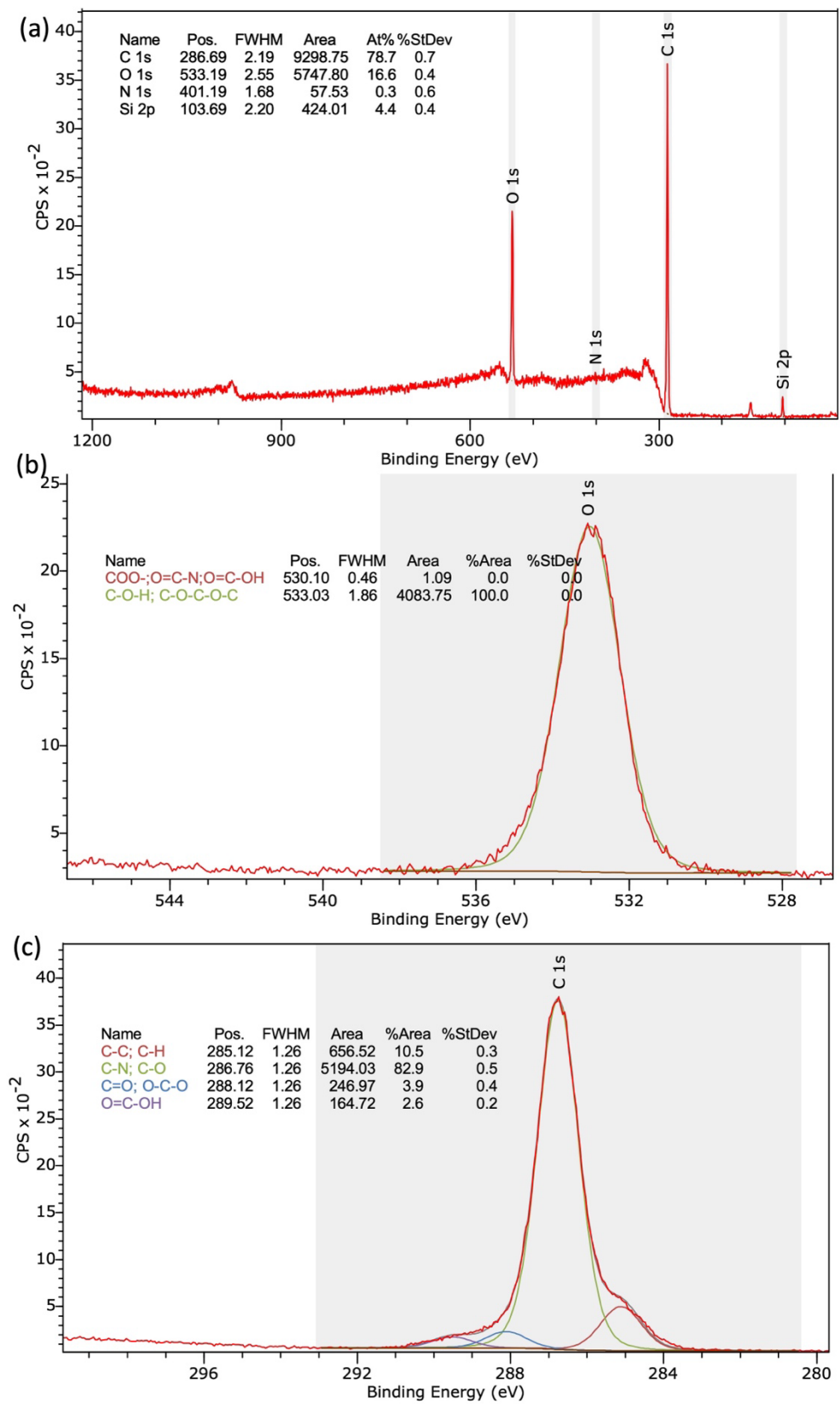

*Figure S3. Characterisation of the chemical gradient across the diamond surface treated with air plasma: Position 1 (less exposed side) (a) shows the entire XPS spectrum (b) an analysis of the oxygen group peak. (c) is the area of the C1s peak*

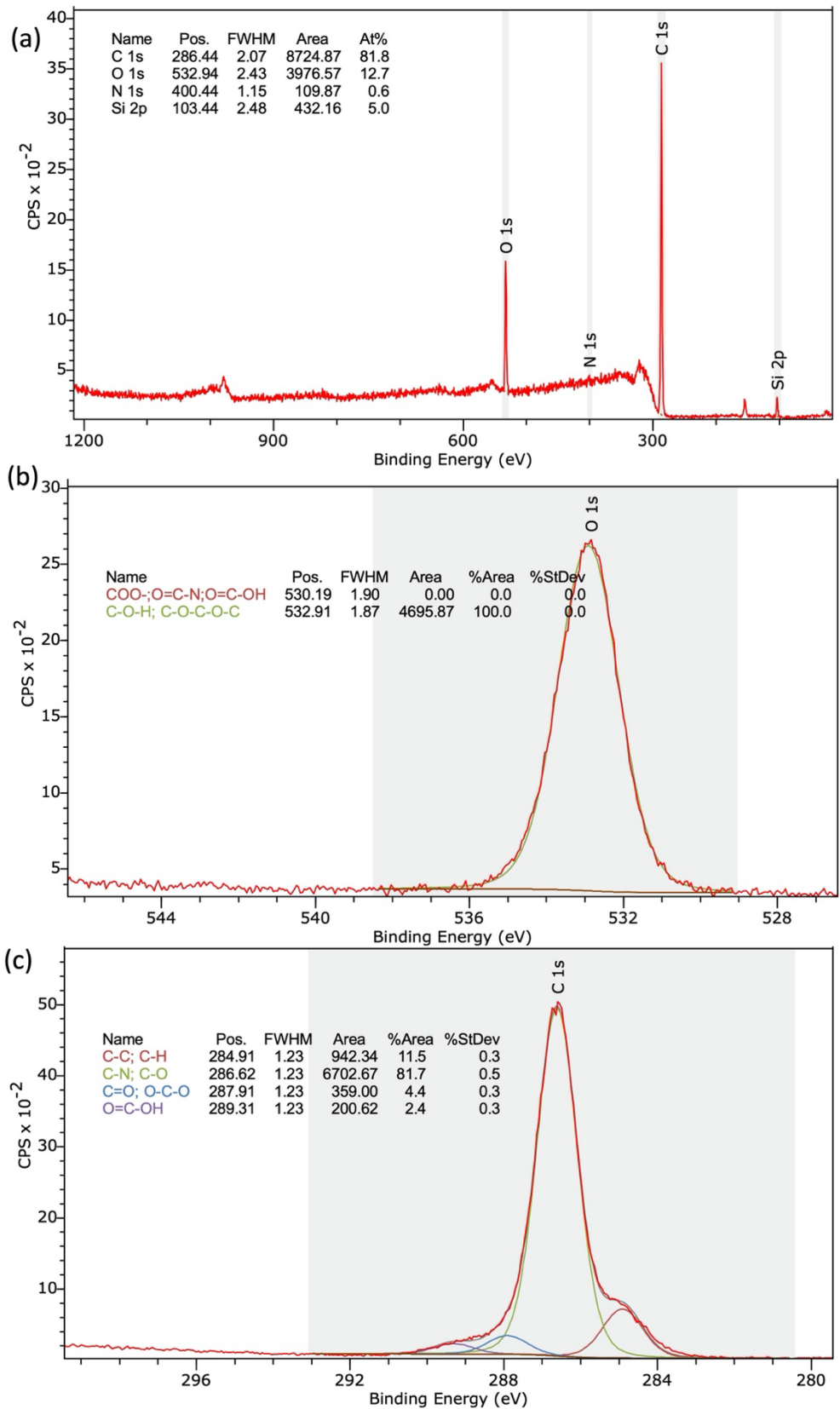

*Figure S4. Characterisation of the chemical gradient across the diamond surface treated with air plasma: Position 2 (more exposed side) (a) shows the entire XPS spectrum (b) an analysis of the oxygen group peak. (c) is the area of the C1s peak*

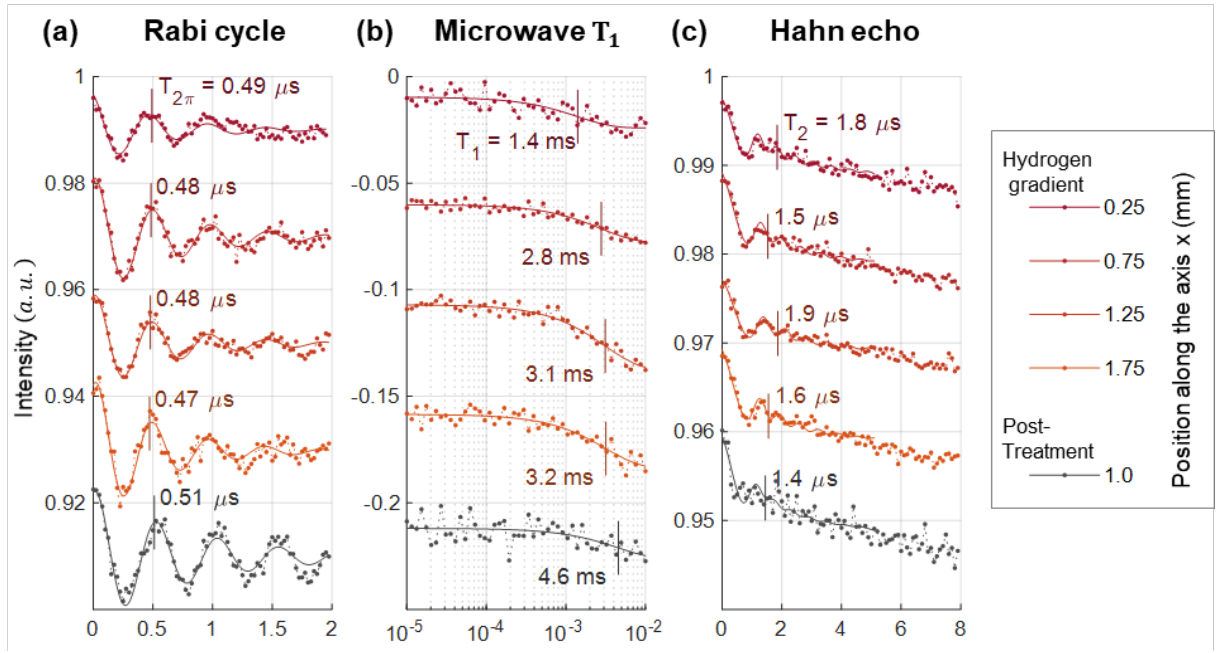

*Figure S5. Rabi oscillations (a) T1 microwaves (b) and Hahn echo sequence performed in confocal configuration.*

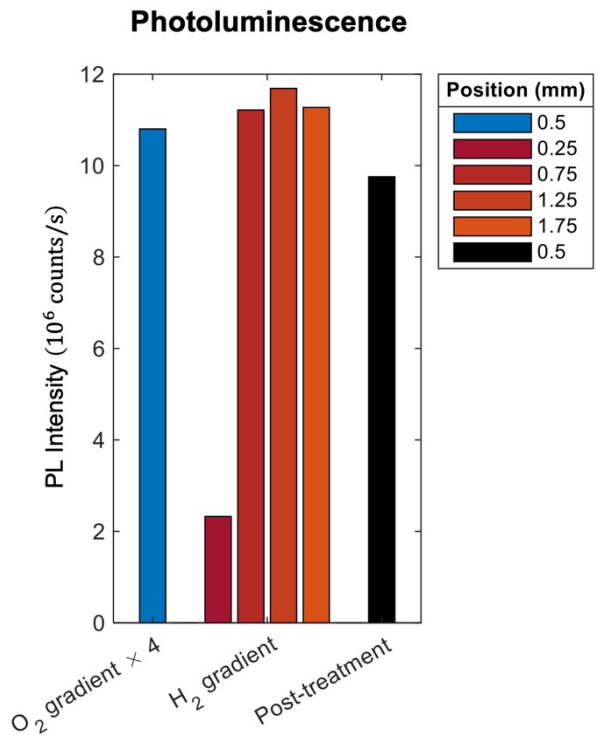

*Figure S6: Photoluminensence counts from the surfaces before (blue) directly after at different positions (red) and after regeneration (black). Except for the most exposed side (0.25 mm from the open side of the mask), the count rates are relatively constant.*

---

[1] Ghodbane, S., Ballutaud, D., Omnès, F. and Agnès, C., 2010. Comparison of the XPS spectra from homoepitaxial {111},{100} and polycrystalline boron-doped diamond films. *Diamond and related materials*, *19*(5-6), pp.630-636.

[2] Yang, L., Jiang, C., Guo, S., Zhang, L., Gao, J., Peng, J., Hu, T. and Wang, L., 2016. Novel diamond films synthesis strategy: methanol and argon atmosphere by microwave plasma CVD method without hydrogen. *Nanoscale Research Letters*, *11*, pp.1-6.